\documentclass[11pt]{article}
\usepackage[margin=1in]{geometry}
\usepackage{graphicx}
\usepackage{amssymb}
\usepackage{algorithm}
\usepackage{algorithmic}
\usepackage{epstopdf}
\usepackage{url}
\usepackage{amsmath,bm}
\usepackage{subfigure}
\usepackage{multirow}
\usepackage{mdwlist}
\usepackage{authblk}
\usepackage{times}
\usepackage{wrapfig}
\usepackage{cite}
\usepackage[compact]{titlesec} 
\usepackage[colorlinks,urlcolor=blue]{hyperref}
\begin{document}     
	\begin{center}
		\Large Website Detection Using Remote Traffic Analysis
\vskip 1.0ex		
		\large Xun Gong$^\dagger$ \qquad Negar Kiyavash$^\ddagger$ \qquad Nab{\'i}l Schear$^*$ \qquad Nikita Borisov$^\dagger$ 
\vskip 1.0ex
		\normalsize
		$^\dagger$Department of Electrical and Computer Engineering \\
		$^\ddagger$Department of Industrial and Enterprise Systems Engineering \\
		$^*$Department of Computer Science \\
		University of Illinois at Urbana-Champaign  \\
		\{\texttt{xungong1},\texttt{kiyavash},\texttt{nschear2},\texttt{nikita}\}\texttt{@illinois.edu}
	\end{center}

\begin{abstract}

Recent work in traffic analysis has shown that traffic patterns leaked through side channels can be used to recover important semantic information. For instance, attackers can find out which website, or which page on a website, a user is accessing simply by monitoring the packet size distribution.  
We show that traffic analysis is even a greater threat to privacy than previously thought by introducing a new attack that can be carried out remotely.  In particular, we show that, to perform traffic analysis, adversaries do not need to directly observe the traffic patterns. Instead, they can gain sufficient information by sending probes from a far-off vantage point that exploits a queuing side channel in routers.

To demonstrate the 	threat of such remote traffic analysis, we study a remote website detection attack that works against home broadband users. Because the remotely observed traffic patterns are more noisy than those obtained using previous schemes based on direct local traffic monitoring, we take a dynamic time warping (DTW) based approach to detecting fingerprints from the same website.  As a new twist on website fingerprinting, we consider a website detection attack, where the attacker aims to find out whether a user browses a particular web site,
and its privacy implications.
We show experimentally that, although the success of the attack is highly variable, depending on the target site, for some sites very low error rates.  We also show how such website detection can be used to deanonymize message board users.

\end{abstract}

\section{Introduction}

Traffic analysis is the practice of inferring sensitive information from patterns of communication.  Recent research has shown that traffic analysis applied to network communications can be used to compromise users' secrecy and privacy.
By using packet sizes, timings, and counts, it is possible to fingerprint websites visited over an encrypted tunnel~\cite{bissias+:pet05, liberatore-levine:ccs06, chen+:oakland10,Herrmann2009}, infer keystrokes sent over a secure interactive connection~\cite{song+:sec01,zhang-wang:sec09} and even detect phrases in VoIP sessions~\cite{wright+:oakland08,whitephonotactic,wright+:tissec10}.  These attacks have been explored in the context of a \emph{local adversary} who can observe the target traffic directly on a shared network link or can monitor a wireless network from a nearby vantage point~\cite{kohno-devices-that-tell-on-yo}.

We consider an alternate traffic analysis approach that is available to \emph{remote} adversaries.  We notice that it is possible to infer the state of a router's queue through the observed queueing delay of a probe packet.  By sending frequent probes, the attacker can measure the dynamics of the queue and thus learn an approximation of the sizes, timings, and counts of packets arriving at the router.  In the case of home broadband networks, in particular, DSL lines, the attacker can send probe packets from a geographically distant vantage point, located as far away as another country; the large gap between the bandwidth of the DSL line and the rest of the Internet path makes it possible to isolate the queueing delay of the ``last-mile'' hop from that experienced elsewhere.

To demonstrate the feasibility of using remote traffic analysis for real malicious attack to learn sensitive information, we adapt the website fingerprinting attack~\cite{bissias+:pet05,liberatore-levine:ccs06,Herrmann2009}, previously targeted at local victims, to a new scenario and introduce a remote website detection attack.
Our attack can find out when a victim user under observation visits a particular target site without directly monitoring the user's traffic. 
This would allow, for example, a company to find out when its employees visit its competitors' sites \emph{from their home computers} or 
deanonymize users of web boards. 

In our adaptation, we encountered two challenges: the information obtained through remote traffic analysis is more noisy than in the local case, and there is no easily available training set from which to create a fingerprint.  To address the former problem, we improved on the previous inference methodology, which used the distribution of packet sizes and inter-arrival times, and developed a fingerprint detection technique that makes use of ordered packet size sequences and the dynamic time warping (DTW) distance metric.  To create a training set, we designed a testbed that uses an emulated DSL link and a virtual execution environment to replicate the victim's home environment.

To evaluate our work, we sent probes to a home DSL line in the United States from a rented server in a data center near Montreal, Canada; we chose this set up to demonstrate the low cost and barrier to entry to conduct the attack.  We then compared the probe results with profiles of website fetches generated in a virtual testbed at our university. 
We tested our attack on detecting each of a list of 1\,000 popular websites.  We found that detection performance was highly variable; however, for a significant fraction of sites, it was possible to obtain very low false-positive rates without incurring significant false negatives.  We also found that there is some accuracy loss due to the discrepancies in the test and training environments (distant from each other) that we were not (yet) able to eliminate.  If the  training and test data are both collected from the same location, a much larger
fraction of sites can be accurately detected with low error rates.  We find that despite working with a much noisier information source than previous web fingerprinting work~\cite{bissias+:pet05,liberatore-levine:ccs06,Herrmann2009}, our website detection attack nevertheless shows that remote traffic analysis is a serious threat to Internet privacy.

The rest of the paper is organized as follows.  We describe our approach to remote traffic analysis in~\S\ref{sec:remote}.  In~\S\ref{sec:finger} we describe our adaptation of previous website fingerprinting attack to remotely confirming user's browsing activities. 
We evaluate our website detection attack  in~\S\ref{sec:eva}.  We then discuss further extensions and the limitations of our technique in~\S\ref{sec:discussion} and present related work in~\S\ref{sec:related}, concluding in~\S\ref{sec:con}.

\section{Remote Traffic Analysis}
\label{sec:remote}

\begin{figure}[t]
 \begin{center}
    \includegraphics[width=0.6\columnwidth]{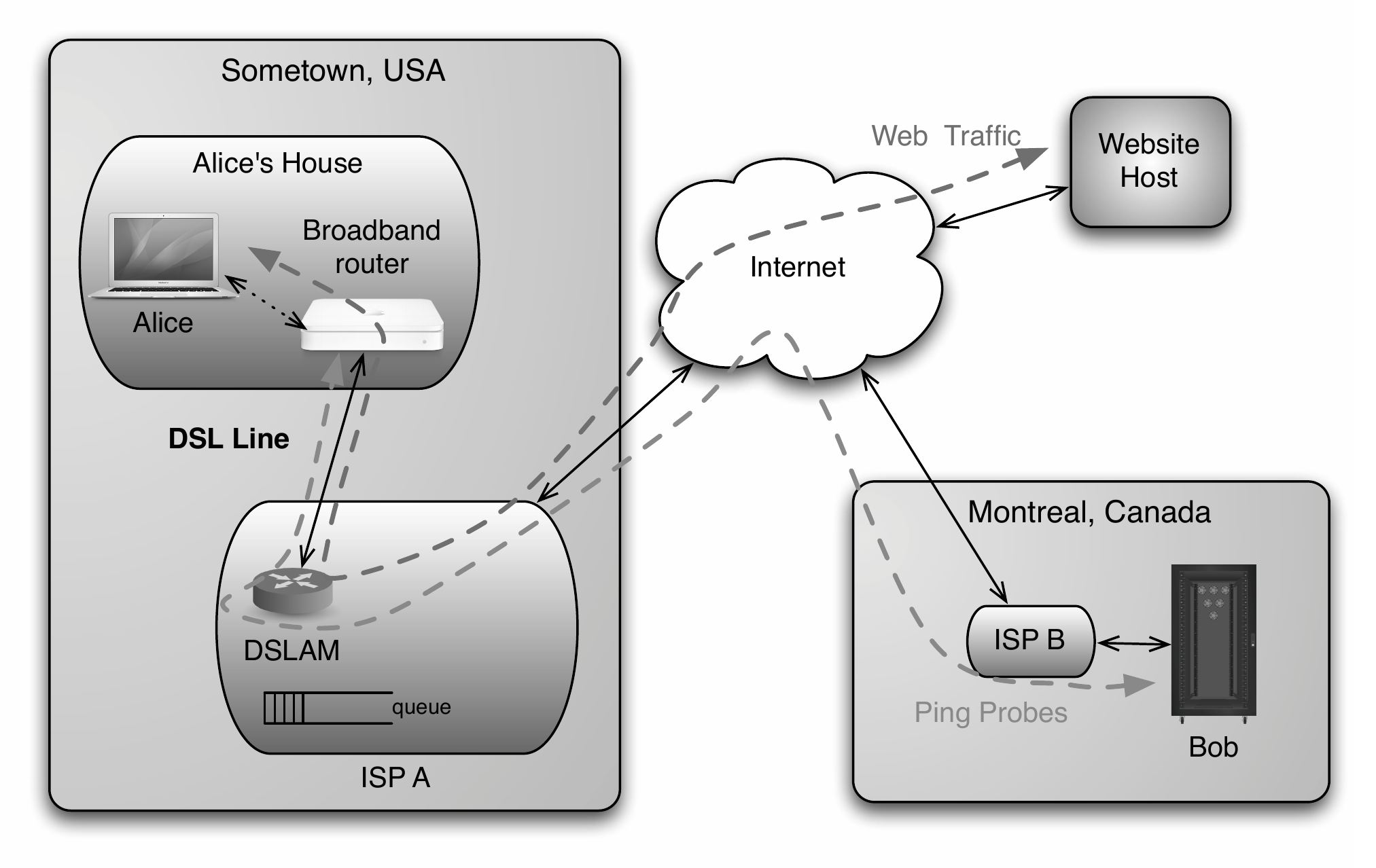}
    \caption{Queueing side channel}
    \label{fig:diagram}
   \end{center}
\end{figure}

Traffic analysis attacks have been known to be effective for quite some time.
And yet, for most Internet users, they represent a minor concern at best.
Although a dedicated attacker could always intercept traffic by, say,
bribing a rogue ISP employee, or tapping a switch box, he would run
the risk of being caught and potentially incurring criminal charges.
In any case, this level of effort seems justified only for highly sensitive
material, rather than casual snooping; therefore, as long as sensitive
data are protected by encryption or other techniques, a user may feel
relatively safe.

We show, however, that traffic analysis can be carried out at a significantly
lower cost, and by attackers who never come into physical proximity with
the user.  In fact, the attackers can launch their attacks from another state or
country, as long as they have access to a well-provisioned Internet connection.
This, in turn, is very easy to obtain due to the highly-competitive Internet hosting
business sector: a virtual private server in a data center can cost as little as a few dollars a month.\footnote{See, for example, \url{www.vpslink.com} (retrieved February 2011).}
We show that the attacker's traffic is very low rate, thus attackers do not need to incur high bandwidth
costs.  Furthermore, users who are being spied upon are unlikely to
notice the small amount of performance overhead.  Thus, anyone with a credit card\footnote{Working stolen credit cards are an easily acquired commodity on the black market~\cite{franklin+:ccs07}.}
can carry out the attack and leave little trace.
  
\begin{wrapfigure}{r}{0.5\textwidth}
   \centering
    \subfigure[Alice's traffic pattern]{
    \label{fig:yahoo}
    \includegraphics[width=0.5\columnwidth]{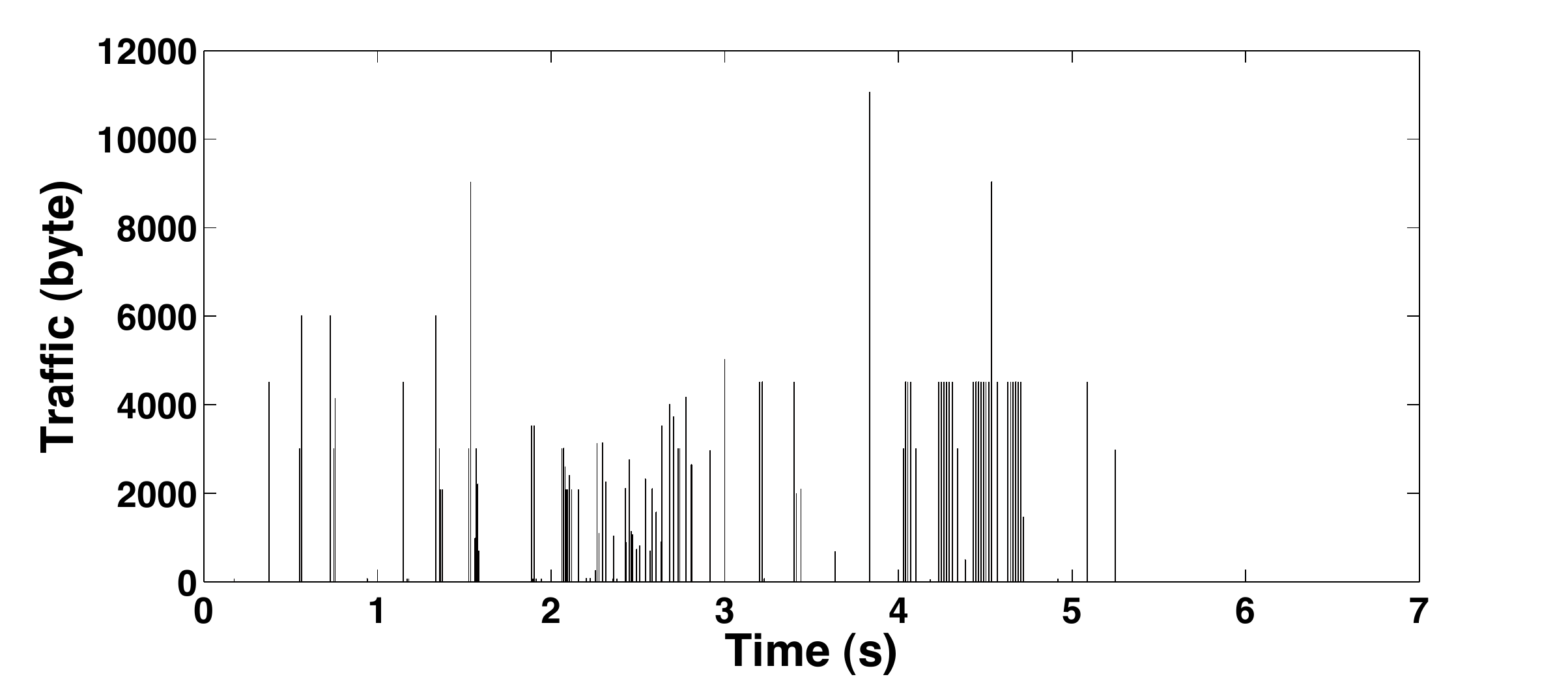}
      } 
	\subfigure[RTTs measured by Bob]{
       \label{fig:rtt}
    \includegraphics[width=0.5\columnwidth]{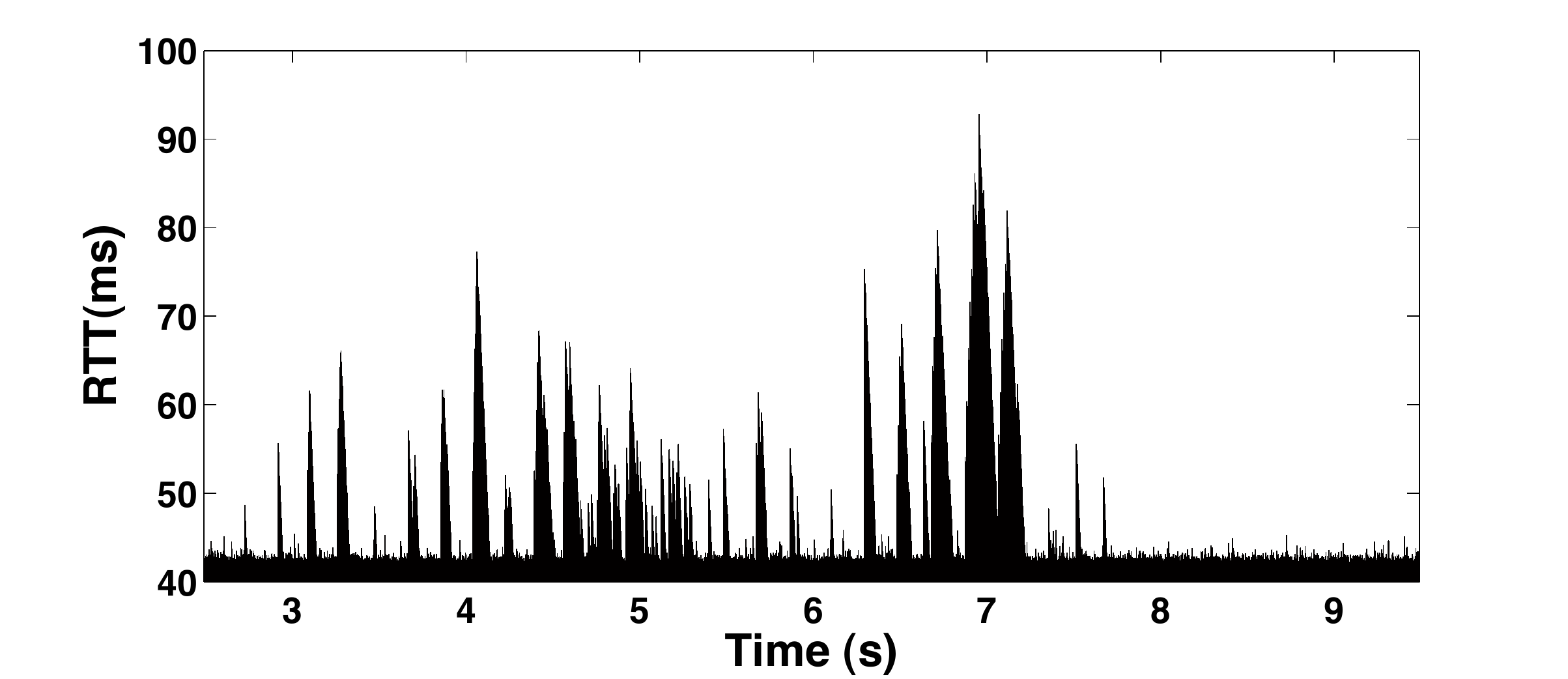}
      } 
        \subfigure[Recovered traffic pattern]{
       \label{fig:rtt-processed}
    \includegraphics[width=0.5\columnwidth]{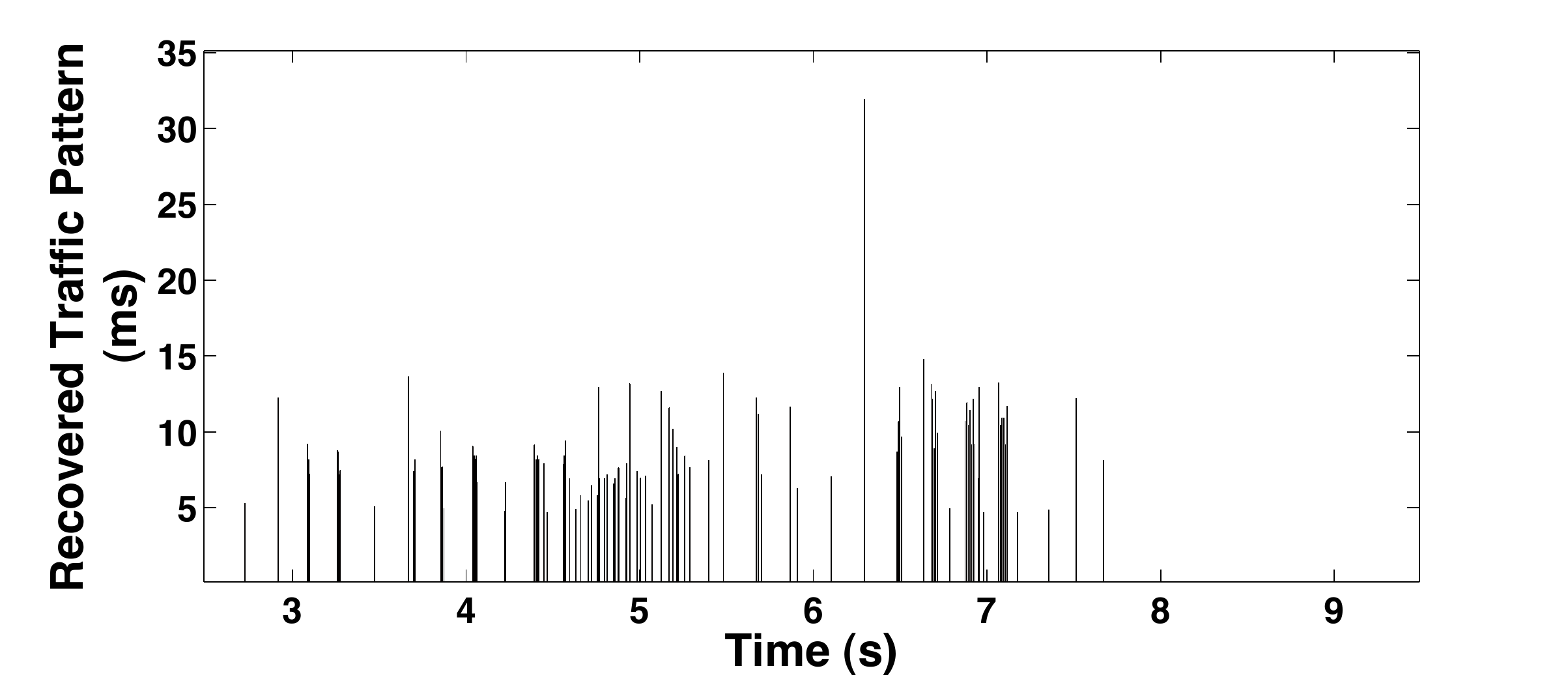}
      }
    \caption{Real traffic on a DSL vs. probe RTTs. Alice resides in some town in US. Bob is located in Montereal, Canada.}
    \label{fig:trafficvsrtt}
\end{wrapfigure}
In this section, we describe our approach to remote traffic analysis. We first introduce the queueing side channel, which is the basis of the attack. Then we design an algorithm to recover users' traffic patterns from the information leaked through this side channel.

\subsection{Queuing Side Channel}
\label{subsec:QSC}

We consider the following scenario. Alice is a home user at Sometown USA, browsing a website via her DSL Internet connection.  Her computer is connected to a broadband router, using a wireless or wired LAN connection.\footnote{In some cases, Alice's computer might be connected to the DSL line directly.}  The router is connected via a DSL line to a DSLAM\footnote{DSL access multiplexer} or similar device operated by her ISP, which is then (eventually) connected to the Internet.
Unbeknownst to Alice, Bob, who is located in another state, or another country wishes to attack Alice's privacy.  If Bob knows Alice's IP address (for example, if Alice visited a site hosted by Bob), he can use his computer to send a series of ICMP echo requests (pings) to the router in Alice's house and monitor the responses to compute the round-trip times (RTTs).  One component of the RTTs is the queueing delay that the packets experience at the DSLAM prior to being transmitted over the DSL line; thus the RTTs leak information about the DSLAM queue size.  This leakage in turn reveals traffic patterns pertaining to Alice's activities.

Since the probe packets traverse many Internet links, and the queuing delays on Alice's DSL link are but one component of the RTT, the question is, how much information is leaked by this side channel? Furthermore, can it be used to infer any information about Alice's activities? To evaluate the potential of this attack, we carried out a test on a home DSL link located in the USA. In the test, Alice opens a Web page \href{http://www.yahoo.com/}{www.yahoo.com} on her computer. Simultaneously, Bob in Canada sends a ping request every 10\,ms  to Alice's home router. Figure~\ref{fig:yahoo} depicts the traffic pattern of Alice's download traffic. The height of each peak in the figure represents the total size of packets that are downloaded during each  10\,ms interval. Figure~\ref{fig:rtt} plots the RTTs of Bob's ping requests. We can see a visual correlation between the traffic pattern and observed RTTs; whenever there is a large peak in the user's traffic, the attacker observes a correspondingly large RTT. 

The correlation between Alice's traffic and Bob's observed probe RTTs can be explained as follows. 
The RTTs include both the queuing delay incurred on the DSL link and delays on intermediate routers, which sit between Bob's computer and Alice's router.
The intermediate routers are typically well provisioned and are unlikely to experience congestion~\cite{Lakshminarayanan2003,akella+:imc03}; furthermore,
the intermediate links have high bandwidth and thus queueing delays will be small in all cases.  We validate this using our
own measurements in the next subsection.

On the other hand, Alice's DSL link is, by far, the slowest link that both her traffic and Bob's probe are likely to traverse. The queues at Alice's router can grow to be quite long (in relative terms), due to TCP behaviors, which cause the \url{www.yahoo.com} server to send a batch of TCP packets at a fast rate.  As most  routers schedule packets in a First In First Out (FIFO) manner, this congestion will leads to large queuing delays of Bob's ping packets. 
We saw that the additional delay caused by Alice's incoming traffic could be as high as over 100\,ms. Thus, Alice's traffic patterns are clearly visible in the RTTs.

\begin{figure}[t]
   \centering
 
 \subfigure[Empirical CDF of ping RTTs from a typical host.]{
   \label{fig:raw-rtt}
     \includegraphics[width=0.5\columnwidth]{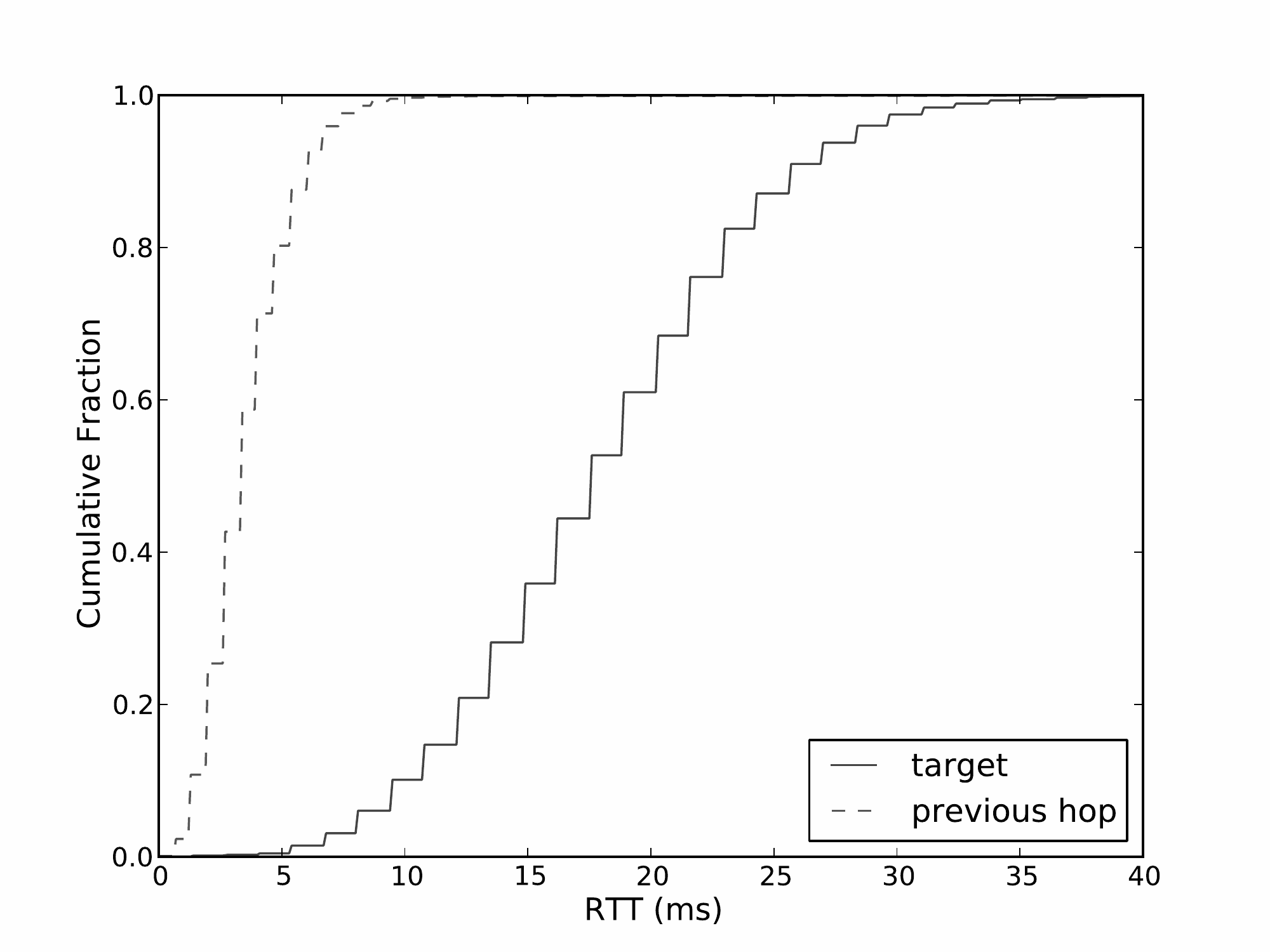}
      }\subfigure[Empirical CDF of 95th percentile minus minimum RTTs. ]{
       \label{fig:full-cdf}
    \includegraphics[width=0.5\columnwidth]{zoom-95.pdf}
      }
         \caption{Measurement of DSL Probe Variance}
     \end{figure}

\begin{table}[t]
\centering
\begin{small}
\caption{RTTs measured by pinging from worldwide advantage points (in {\bf ms})}
\begin{tabular}{|c|c|c|c|}\hline
 
{ Node} & { Mean} &{ StdDev}  \\ \hline \hline

\texttt{planet6.cs.ucsb.edu}           &        66.590          &         0.543\\ \hline

\texttt{pl1.rcc.uottawa.ca}          &       42.936           &         0.619\\ \hline

\texttt{pl1.grid.kiae.ru}          &           153.205            &     0.749 \\ \hline

\texttt{planetlab2.c3sl.ufpr.br}  &      177.318           &      0.868 \\ \hline

\texttt{planet2.pnl.nitech.ac.jp} &       197.043      &           0.567 \\ \hline
 \texttt{pl1.eng.monash.edu.au}    &        221.460     &           1.784 \\ \hline
\texttt{planetlab1.xeno.cl.cam.ac.uk}  & 319.752     &           2.297 \\ \hline 

\texttt{planetlab2.comp.nus.edu.sg}  &    291.193     &          4.221 \\\hline
%
\end{tabular}
\label{tab:rtt}
\end{small}
\end{table}

\subsection{Measurement of DSL Probe Variance}

To further confirm that the fluctuation of user's RTTs are primarily determined by the congestion at the DSL link, we conducted a small Internet measurement study.  We harvested IP addresses from access logs of servers run by the authors.  We noted that many DSL providers assign a DNS name containing ``dsl" to customers.  Using reverse DNS lookups, we were able to locate 918 potential DSL hosts.  To determine each host's suitability for measurement, we first determine if it responds to ping requests.  We then use traceroute to locate the hop
before the target DSL host (e.g., the DSLAM).  Next, we ensure this previous hop also responds to ping requests.  Lastly, we measure the minimum RTT of several hundred ping probes.  We exclude any host with a minimum RTT of greater than 100ms  to bound the study to hosts in a wide geographical area around our Montreal, Canada probe server.  Using this method, we identified 189 DSL hosts to measure.  The measurement consists of sending ping probes every 2\,ms for 30 seconds to the target DSL host and then to its previous hop.  We collected these traces in a loop over a period of several hours.

We found that, on average, the target host RTT was $\sim$10\,ms greater than the previous hop.  We frequently observed the pattern in Figure~\ref{fig:raw-rtt} where the previous hop RTT was very stable and target RTT variations greater than 10\,ms.  We then measured the span between the 95th percentile and the minimum observation in each sample.  Figure~\ref{fig:full-cdf} shows the CDFs of this data for the each target DSL host and its previous hop from the measurement set.  We observe that the previous hop RTT span shows more stability than the end host, confirming the one of the primary assumptions of our work. 

To confirm the feasibility of learning the target user's traffic remotely (from another city, or even another country), we examined the RTTs from advantage points across different locations. In this experiment, we used a series of PlanetLab nodes located in 8 countries to send frequent ping requests to a DSL IP address in US, while keeping the host behind the DSL idle; i.e., no background traffic went through the DSL link. The measured  RTT standard deviations, as listed in Table~\ref{tab:rtt}, thus represented the background noises for learning target traffic patterns. Recall in the queuing side channel,  the user's traffic patterns are leaked through the extra delays experienced by the probe ping packets. On a common home DSL link with download speed of 3 Mbps, the delay induced by one packet with size of 1500 bytes is about 4\,ms. This is much higher than the standard deviations of RTTs at the first 5 nodes. Therefore, the traffic patterns would survive well in the RTTs observed on those nodes. 
In fact, many of the patterns we observe involve multiple back-to-back packets, creating extra delay of tens or even hundreds of milliseconds.
Such patterns would be detectable even on the worst of these links, although for best quality of attack, the attacker should pick a vantage point relatively close to the target (e.g., same continent).


\subsection{Traffic Pattern Recovery Algorithm}
\label{subsec:recover}

\begin{wrapfigure}{r}{0.5\textwidth}
\centering
    \includegraphics[width=0.5\textwidth]{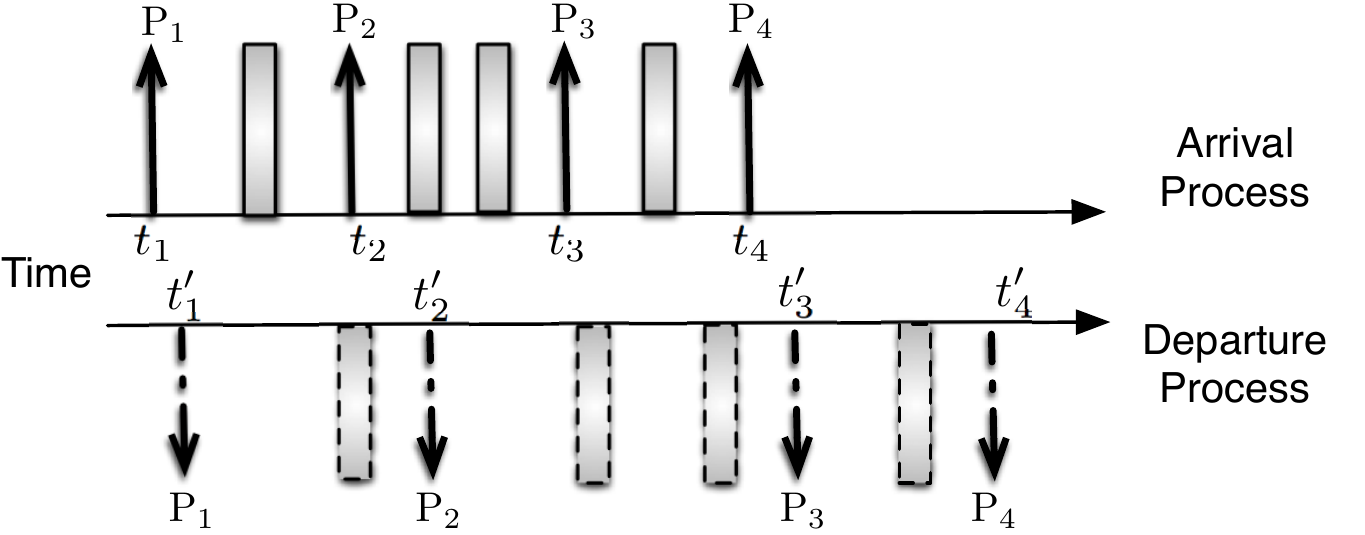}
         \caption{FIFO queuing in the DSL router}
    \label{fig:fifo}
\end{wrapfigure}

We now show how the attacker can analyze the information leaked through this queueing side channel. 
We model the incoming DSL link as a FIFO queue. As most traffic volume in an HTTP session occurs on the download side, we will ignore the queuing behavior on the outgoing DSL link, though it could be modeled in a similar fashion.

Figure~\ref{fig:fifo} depicts the arrival and departure process in this queuing system.  The arrows are Bob's ping packets, denoted by P$_i$'s, and the blocks represent HTTP packets downloaded by Alice. 
The DSLAM serves packets in FIFO manner and at a constant service rate; i.e., the service time is proportional to the packet size. 
As most HTTP packets are more than an order of magnitude larger than ping packets, we ignore the service time for pings.

Assume that ping packet P$_i$ arrives in the queue at time $t_i$, waits for the router to serve all the packets currently in the router, and then departs at time $t'_i$.  Let us consider the observed RTT of the ping packet P$_i$; we can represent it as:
\begin{equation}
	RTT_i = \sum_{l \in \text{links on path}} q^l_i + t^l_i + p^l_i
\end{equation}
\noindent where $q^l_i$, $p^l_i$, and $t^l_i$ are the queueing, propagation, and transmission delays incurred by packet P$_i$ on link $l$.  Note that 
the propagation and transmission delays are mostly constant, and in fact we can approximate:
\begin{equation}
	\sum_{l \in \text{links on path}} t^l_i + p^l_i \approx \min_j RTT_j
\end{equation}
\noindent since Bob is likely to experience near-zero queueing delays for some of the pings.
Furthermore, as argued in~\S\ref{subsec:QSC}, the queueing delay on links other than the DSL line are going to be minimal, thus we can further approximate:
\begin{equation}
	RTT_i \approx \min_j RTT_j + (t'_i - t_i) \label{eq:1}
\end{equation}
Making use of the queuing delay $t'_i-t_i$ from  \eqref{eq:1}, the attacker Bob can further infer the total size of HTTP packets arriving during the interval $[t_{i-1},t_i]$'s, which produces a similar pattern as Alice's traffic in Figure~\ref{fig:yahoo}. For this purpose, two cases need to be considered.

\begin{enumerate*}
\item  $t_i \geq t'_{i-1}$.
In this case, when P$_i$ enters the queue, the DSLAM is either idle or serving packets destined for Alice. 
The delay $t'_i-t_i$ reflects the time required to finish serving the HTTP packets currently in the buffer, and is thus approximately proportional to the total size of Alice's arrivals during the interval $[t_{i-1},t_i]$.
P$_2$ in Figure~\ref{fig:fifo} is one example of this case. 
\item  $t_i<t'_{i-1}$.
In this case, P$_{i-1}$ is still in the queue when P$_i$ arrives. Only after P$_{i-1}$ departs at $t'_{i-1}$, the router can start to serve packets that arrived in the interval $[t_{i-1},t_i]$.  
Thus the delay $t'_i-t'_{i-1}$ is the service time for those packets and can be used to recover the total size. P$_4$ in Figure~\ref{fig:fifo} is one example of this case. 

\end{enumerate*}

\begin{algorithm}[t]
\caption{Traffic pattern recovery algorithm}
\label{algo:tra}
\begin{algorithmic}[1]
\STATE  \textbf{let} \ensuremath{t'_0=0}
\FOR {$i =1$ to \emph{the probe sequence length}}

\STATE \# reconstruct the arrival and departure times 
\STATE  $t_i=    t_{ping}\cdot i$
\STATE $t'_i=RTT_i-RTT_{min}+t_i$

\STATE  \# estimate the total size of packets arriving in  $[t_{i-1},t_i]$
\STATE $\widehat{s_i}= t'_i-max(t'_{i-1},t_i)$

\STATE \# discard noise
\IF {$\widehat{s_i}<\eta$} 
\STATE $\widehat{s_i}=0$
\ENDIF
\ENDFOR
\end{algorithmic}
\end{algorithm}

Algorithm~\ref{algo:tra} summarizes the traffic pattern recovery procedure based on these observations.  To account for minor queueing delays experienced on other links, we define a threshold $\eta$ such that RTT variations smaller than $\eta$ are considered noise and do not correspond to any packet arrival at the DSLAM. 
Figure~\ref{fig:rtt-processed} plots the pattern extracted from RTTs in Figure~\ref{fig:rtt}.  After processing, the resulting time series proportionally approximate the packet size sequence of the original traffic in Figure~\ref{fig:yahoo}. As will be shown in the next section, it can be applied to infer more information about Alice's activities, e.g., website fingerprinting.

Note that in case 1, the attacker may underestimate the size of the HTTP packets arriving in the period $[t_{i-1},t_i]$ because a portion of them will have 
already been serviced by time $t_i$.  The error depends both on the frequency of the probes and the bandwidth of the DSL link.  Since most HTTP packets are of maximal size (MTU), we can ensure that all such packets are observed by setting the ping period to be less than:$\frac{\text{MTU}}{\text{DSL bandwidth}}$.
Thus the adversary must tune the probe rate based on the DSL bandwidth and faster links will require a higher bandwidth overhead (but the pings will form a constant, small fraction of the overall DSL bandwidth.)

\subsection{Properties}

Our remote traffic analysis attack has the following properties that are different from traditional local traffic analysis techniques:
\begin{description}
\item[Remote vantage point]  The attacker does not need to physically capture the target traffic flow. He can launch this attack from almost anywhere, even different states or counties from the target user.

\item[Low cost]  
The attacker can perform this attack as long as he has access to a well-provisioned Internet connection. Moreover, the probe traffic has a very low rate, e.g., 50\,Kbps for 100 pings per second. Thus the attacker does not need to incur considerable bandwidth costs, and victims are unlikely to notice the additional overhead.  Additionally, due to such low cost, the attacker can possibly monitor \emph{multiple targets} simultaneously.

\item[Coarser observation]
From the queueing side channel, the attacker only can obtain estimations about the sum of packet sizes arriving between successive pings and may miss some of the traffic, depending on the probe frequency. This observation is coarser than previous traffic analysis work, where local vantage points enable the attacker to gather information about every single packet, such as the exact size and inter-packet delays.  Thus the performance of a remote traffic analysis attack will generally be worse than what is possible with local observations.  We show that despite coarse observation we are still able to reconstruct an alarming amount of information from remote hosts using the attack.
\end{description}

\subsection{Feasibility}

We have shown that the attacker can recover the user's traffic pattern through the information leakage of the queuing side channel. 
We now address the feasibility of our attack by further discussing the prevalence of the conditions required for this attack.

\subsubsection{ICMP support} The attack scenario we show above relies on ICMP probe packets, hence we care about whether ICMP is enabled in real routers.
 In testing over 918 probable DSL hosts on the Internet, we found over 25\% responded to ping requests.  Since we harvested these probable DSL hosts from the Internet over a period of several months, it is not clear how many that failed to respond were simply down rather than blocking our probes.  Thus, we can assume that the fraction of hosts
 that respond to ping is even larger.  Additionally, in a brief  survey of consumer-grade router hardware, we found that many of them do not perform ICMP filtering, at least not in the default configuration.  Moreover, even though the ping packets are blocked by firewalls on some home routers, other forms of probes may be exploited as well; for example, if the home router exposes TCP ports for file sharing or other applications, SYN packets can be used as probes with the same effectiveness.

\subsubsection{FIFO scheduling policy}  The high correlation between Alice's traffic pattern and Bob's ping RTTs comes from the fact that the router serves packets in FIFO order. 
Note that most home routers today do not use QoS extensions and schedule packets on a given link in FIFO order.  
Thus, information leaked by these routers can be exploited with remote traffic analysis.  
Certainly, a fair queuing implementation~\cite{wfq} would reduce the impact that cross-traffic would have on the probe sequence and hence reduce the effectiveness of the side channel, but not entirely eliminate it~\cite{Sachin10}.

\subsubsection{Limited Last-hop Bandwidth}

The information leaked through our side channel are the states of the queue length in the router's buffer. 
Hence, to have nontrivial queues built up in the buffer, the broadband link must have limited bandwidth compared to the rest of the links in the path. 
In our experiments, we have used speeds typical of current home broadband speeds---several Mbps, and our scheme worked well in those environments.                      
The deployment of faster links, such as Fiber-to-the-Home (FTTH), 
may reduce the effectiveness of the queueing side channel, but notice
that if the core network is similarly upgraded in speed, the  bandwidth disparity necessary for our attack will remain. 

\subsubsection{Victim's IP address}  In our attack, Bob needs to know Alice's IP address to send the probes.  Although this mapping is typically only explicitly known to ISPs, many protocols, such as file sharing, instant messaging, VoIP, and email, will reveal the IP address of a user. Other forms of IP address reconnaissance may also be possible but are outside the scope of this work.

\section{Website Fingerprinting}
\label{sec:finger}

Previous work on traffic analysis has shown that it is often possible to identify the website that someone is visiting based on traffic timings and packet sizes~\cite{bissias+:pet05,liberatore-levine:ccs06,Herrmann2009}, namely, \emph{website fingerprinting}.  We consider whether it is possible to carry out a similar attack using our remote traffic analysis. We first review the three basic steps in previous work when conducting a website fingerprinting attack. 
\begin{enumerate*}
	\item First, the attacker decides some feature of web traffic used to distinguish websites. 
The feature needs to stay relatively stable for accesses to the same single website, but has significant diversity across different sites. 
For example, Herrmann et al.\ use the size distribution of HTTP packets~\cite{Herrmann2009}.
\item The next step is  the \emph{training} procedure. The attacker needs a training data set of fingerprint samples labeled with corresponding destination websites. Usually, these feature profiles are obtained by the attacker browsing websites himself/herself from the same (or similar) network connection as the  user. 

\item In the final step, the attacker \emph{tests} his knowledge from training on the victim user.  He monitors traffic going to the user and matches extracted features with the profiles in his database.  The one with most similarity is chosen as the website browsed by the user.  
\end{enumerate*}

As compared with  previous work, using our remote traffic analysis technique for identifying websites introduces two additional challenges.
First, previous work used fine-grained information like exact packet size distributions to create features, whereas in our setting this information is not available directly, since the queueing side channel produces only approximate sums of packet sizes.
Second, previous work created a training set from \emph{the same} vantage point that was then used for fingerprinting tests. An attacker performing remote traffic analysis must, of course, use a different environment for collecting the training set, potentially affecting the measured features. 
We describe our approaches to solving these two challenges next.

\subsection{Time Series--Based Feature}
\label{sec:dtw}

Since it is hard to infer information about each single packet from our recovered pattern time series, we use the entire time series, which contains the estimated size of all HTTP packets downloaded during each probe period,  to create one  fingerprint trace. 
Identification of websites is based on the similarity between the observed fingerprints and samples in the training set. 

\begin{wrapfigure}{r}{0.5\textwidth}
  \centering
    \includegraphics[width=0.4\columnwidth]{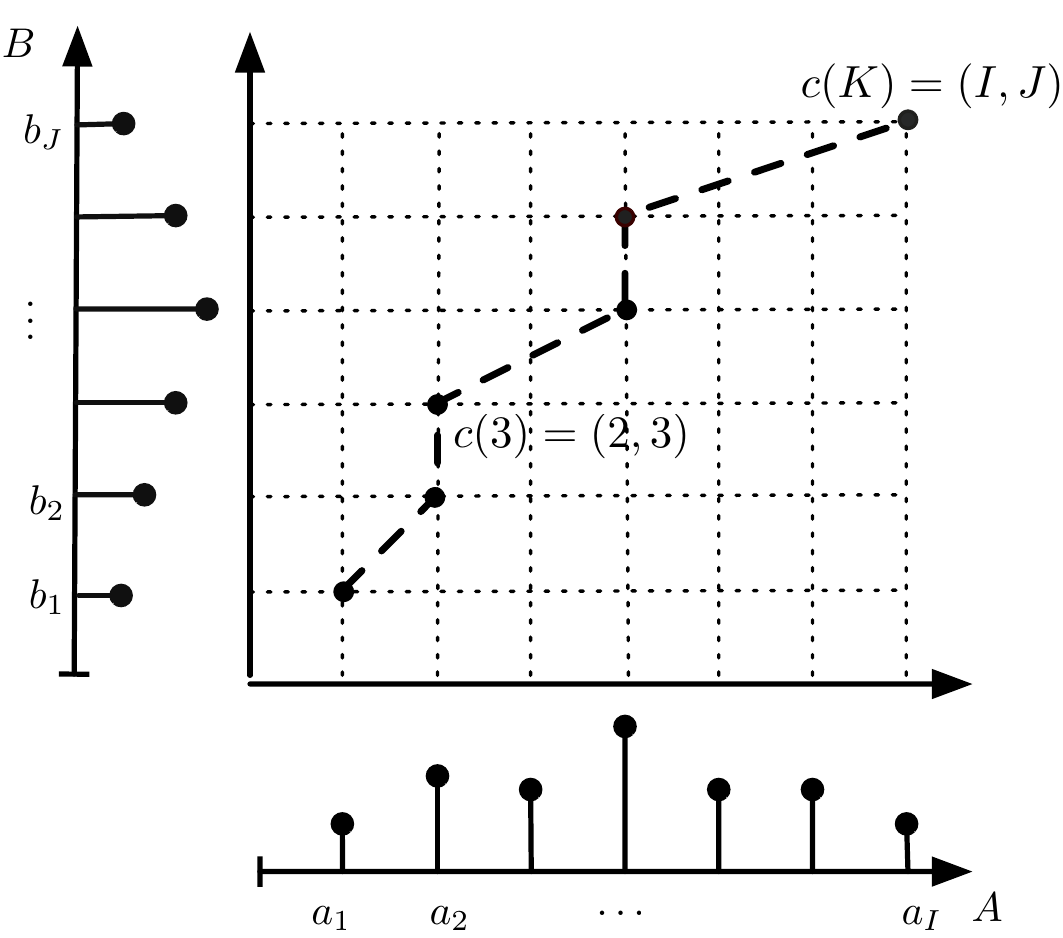}
    \caption{Warping function in DTW}
    \label{fig:dtw}
\end{wrapfigure}

The challenge is to find a meaningful distance between fingerprint traces. 
Note that pointwise comparisons will produce poor results. This is because parts of the fingerprint may be impacted by the noise from a small queueing delay on a core Internet link. 
Additionally, the fingerprint could miss some packets contained in the original traffic due to pattern recovery errors. Finally, even fingerprints of the same website are not strictly synchronized in time due to the inter-packet delay variations. To deal with these issues, we turn to the Dynamic Time Warping (DTW) distance~\cite{Sakoe78}. 
DTW was developed for use in speech processing to account for the fact that when people speak, they pronounce various features of the phonemes
at different speeds, and do not always enunciate all of the features.  DTW attempts to find the best alignment of two time series by creating a non-linear time warp between the sequences. 
Figure~\ref{fig:dtw} visualizes the DTW-based distance between two time series: $A=\{ a_1, a_2\dots, a_I\}$ and $B=\{b_1, b_2\dots, b_J\}$.
Let function $F(c)=\{c(1),\dots,c(K)\}$ be a mapping from series $A$ to series $B$ where $c(k)=(a(i),b(j))$. 
For every pair of matched points based on the mapping, we define the distance as $d(c(k))=d(i,j)=|a_i-b_j|$.
The final distance between the A and B can then be defined as a weighted and normalized sum over all matched point pairs as $D(A,B)=\min_F\left\{\frac{\sum_{k=1}^{K} d(c(k))w(k)}{\sum_{k=1}^{K} w(k)} \right\}\label{dtw_dis}$. The weights $w(k)$'s are flexible parameters picked based on the specific application scenario. Applying dynamic programming, one can find the warping function with minimum distance, which captures the similarity between the two time series under best matched alignment.  

In our attack, we applies DTW-based distance to account for the estimation errors and time desynchronizations in fingerprints. 
Based on the distances with the training data set, the attacker will know if a test sample indicates the activity that the user browsed the website of interest. 


 
 

\subsection{Training Environment}

To obtain an accurate training fingerprint for a particular user's traffic, the attacker must be able to replicate the network conditions on that
user's home network. The approach we use is to set up a virtual machine running a browser that is connected to the Internet
via a virtual Dummynet link~\cite{emulab-dummynet}.  The virtual machine is then scripted to fetch a set of web pages of interest;
at the same time, an outside probe is sent across the Dummynet link, simulating the attack conditions on a real DSL link.  

A number of parameters of the link need to be carefully decided. We found that the most important parameter for the attacker to replicate was the
link bandwidth.  First, as discussed in~\S\ref{subsec:recover}, the probe frequency should be adjusted based on the link bandwidth.
Bandwidth also affects the magnitude of observed queuing delays. Additionally,  it can significantly alter the traffic pattern itself, as TCP congestion control mechanisms are affected by the available bandwidth. Fortunately, estimating the bandwidth on a link is a well-studied
problem~\cite{strauss2003measurement,prasad2003bandwidth,pathchirp}. 
In our tests, we use a packet-train technique by sending a burst of probe packets and measuring the rate at which responses are returned.  
Since most DSL lines have asymmetric bandwidth, we used
TCP ACK packets with 1000 data bytes to measure the download bandwidth on the link.  The target would send a short TCP reset packet for each ACK that it received, with the spacing between resets indicating the downstream bandwidth; we found this method to be fairly accurate.

The round-trip time between the home router and the website hosts also affects the fingerprint. 
When opening a webpage, the browser can download objects from several host servers. 
The traffic pattern is the sum of all download connections, hence the shape of observed fingerprint does depend on the RTTs to these servers. 
However, we did not explicitly model this parameter considering the difficulty to accurately tune up the link delays to multiple destinations.  
The effects to the will be further discussed in~\S\ref{sec:eva}. 


The fingerprint may be affected by the choice of browsers and operating systems as well; for best result, the training environment should model the target as closely as possible.  Information about browser and operating system versions can be easily obtained if the target can be convinced to visit a website run by the attacker; additionally, fingerprinting techniques in~\cite{nmap} may be used to recover some of this information.

\subsection{Attack Scenarios}
\label{sec:attacks}

We consider several attack scenarios that make use of website fingerprinting.  We can first consider the classic website fingerprinting scenario: Bob obtains traces from Alice's computer by sending probes to her DSL router and compares them to fingerprints of websites
that he has generated, in order to learn about her browsing habits.  Note that this can be seen as a \emph{classification} task: each web request in Alice's trace is classified 
as belonging to a set of sites.  This scenario has been used in most of the previous work on website fingerprinting, but it introduces the requirement that Bob must know the set of potential sites that Alice may visit.  Without some prior information about Alice's browsing habits,
this potential set includes every site on the Internet, making it infeasible to generate a comprehensive set of fingerprints.  One could create fingerprints for popular sites only, but this reduces the accuracy of the classification task~\cite{Herrmann2009,Sun02,coull2007web}.
For example, the top 1\,000 US sites,
as tracked by Alexa, are responsible for only 56\% of all page views, therefore, even a perfect classifier trained on the 1\,000 sites
would give the wrong result nearly half the time.\footnote{In fact, the situation is even worse, since Alexa counts \emph{all} page views 
within a certain top-level domain, whereas fingerprints must be created on each individual URL.}

We therefore consider a different scenario, where Bob wants to \emph{detect} whether Alice visits a \emph{particular} site.  For 
example, if Bob is Alice's employer, he may wish to check to see if she is considering going to work for Bob's competitor, Carol.  To carry 
out this attack, Bob would create a fingerprint for Carol's jobs site; he would then perform a binary classification task on Alice's traffic,
trying to decide whether a trace represents a visit to the target site or some other site on the Internet.  As we will see, such binary classification can be performed with relatively high accuracy for some choices of sites.  Note that, as Alice's employer, Bob has plenty of opportunities to learn information about Alice's home network, such as her IP address, browser and operating system versions, and download bandwidth, by observing Alice when she connects to a password-protected Intranet site, and can therefore use this information to create 
accurate training data for building fingerprints.

As another example, Bob may be trying to identify an employee who makes posts to a web message board critical of Bob.\footnote{This example is motivated by several actual cases of companies seeking to do this; see \url{https://www.eff.org/cases/usa-technologies-v-stokklerk} and \url{https://www.eff.org/cases/first-cash-v-john-doe}.}  Bob can similarly build profiles, tailored for each employee's home computer, of the web board and perform remote traffic analysis.  He can then correlate any detected matches to the times of the posts by the offending pseudonym; note that this \emph{deanonymization} attack is able to tolerate a significant number of false-positive and false-negative errors by combining observations over many days to improve confidence~\cite{danezis2005statistical}.

\section{Evaluation}
\label{sec:eva}

We next present our results of website detection attack. First, we describe the experimental setups and data collection procedure. 

 \subsection{Experimental Setups}

\begin{figure}
   \centering
\subfigure[The DSL-Setup for test]{
       \label{fig:dsl}
    \includegraphics[width=0.5\columnwidth]{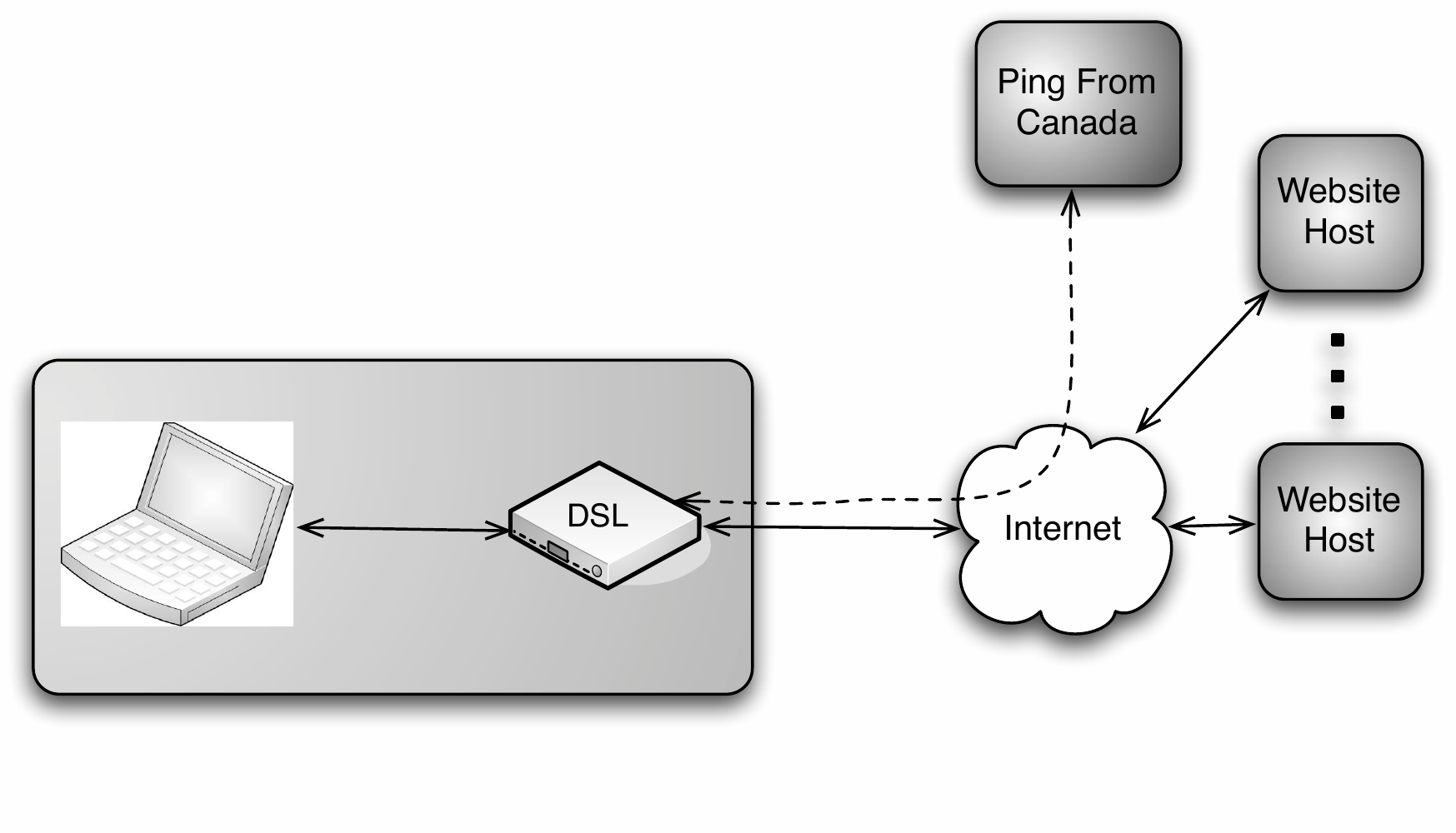}
      }\subfigure[The VM-Setup for training]{
    \label{fig:vm}
    \includegraphics[width=0.5\columnwidth]{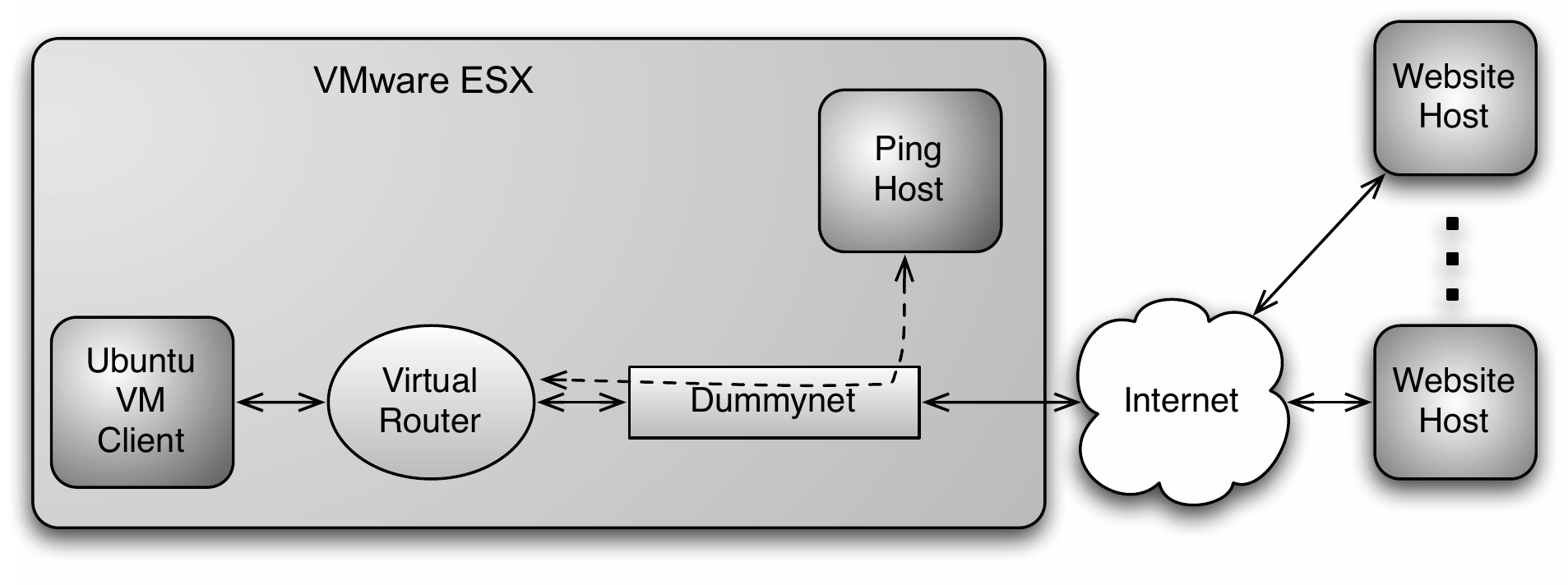}
      }  
       \caption{Experimental setups for website detection}
    \label{fig:setup}
\end{figure}

We built a \emph{DSL-Setup} consisting of a target system and a ping server,  as shown in Figure~\ref{fig:dsl}. The target system captured the real environment of a home user. It ran on a laptop, located in our city (inside US), connected to DSL line with 3\,Mbps download and 512\,Kbps upload speeds. On the laptop, we used a shell script to automatically load websites using Firefox 4.0\footnote{\url{http://www.mozilla.com/firefox/}}. 
The ping server was  a commercial hosting system, located in the Canadian province of Quebec, acting as the remote attacker. 
It was scripted to send pings at  precise time intervals with  hping\footnote{\url{http://www.hping.org}} and record ping traces with tcpdump\footnote{\url{http://www.tcpdump.org}}. We set the ping interval to 2\,ms.

To emulate the attacker's training procedure, we also built a \emph{VM-Setup}, a VMware ESX host testbed located in our lab, as shown  in Figure~\ref{fig:vm}.
On this machine, we ran several VMware guest operating systems:  a Ubuntu VM Client, a virtual router and a host implementing a transparent Dummynet link.
The Ubuntu VM Client acted as a virtual target, and was scripted to browse websites using Firefox, similar to the real home user.
The virtual router provided NAT service for the client, and was connected to the Internet through the Dummynet link.
The Dummynet bridge was configured to replicate the network conditions of the target DSL link (i.e., the bandwidths). 
As in the DSL-Setup, we sent probes from another host outside the constrained Dummynet link to the virtual NAT router periodically. 
The attacker then collected training fingerprints while the virtual client was browsing websites through this virtual `DSL' link. 
Note the virtual router and ping host were connected to the same dedicated high-speed LAN minimizing the impact of 
additional noise added by intermediate routers or network congestion caused by other hosts.

\subsection{Data Collection}

We collected fingerprints of the front pages for 1000 websites 
on the top list on Alexa\footnote{\url{http://www.alexa.com}}. For websites which have multiple mirrors in different countries like \href{http://www.google.com}{www.google.com}, we only considered the site with the highest rank.  
We excluded websites with extremely large loading time (greater than 60\,s). 
For each website, we collected 12 fingerprint samples from both the DSL and VM setups. The delay between collecting two samples is half an hour.  
Following the same assumptions in previous papers~\cite{liberatore-levine:ccs06, Herrmann2009}, the browsers were configured appropriately (no caching, no automatic update checks and no unnecessary plugins). This makes our results comparable with previous work.

\subsection{Website Detection}

We first analyze the ability of an attacker to detect whether a user visits a particular site.  To do so, the attacker checks whether
the distance between the user trace and the target web site is smaller than some threshold, and if so, the web site is considered detected.  This is a binary classification task and its performance can be characterized by the rates of false positives---a different
website incorrectly identified as the target--and false negatives--the target website not being identified.  The choice of threshold $t$
creates a tradeoff between the two rates: a smaller threshold will decrease false positives at the expense of false negatives.

To estimate false-positive rate given a particular threshold $\nu$, we fix a target site $t$ and use the 12 samples $T = \{ s_{t,1}, \ldots, s_{t,12}\}$ as the training set.  We use the samples from the other sites as a test set; i.e., $U = \{ s_{i,j} \}$ for $i \neq t, j \in \{1, \ldots, 12\}$.  Given a sample $s_{i,j} \in U$, we calculate the average distance from it to the training samples:

\begin{equation}
 d_t(s_{i,j}) = \frac{\sum_{k=1}^{12} D(s_{i,j},s_{t,k})}{12}
\end{equation}

\noindent where $D(\cdot,\cdot)$ is the DTW-based distance function defined in \S\ref{sec:dtw}.  We then consider every sample $s_{i,j} \in U$
such that $d_t(s_{i,j}) < \nu$ to be a false positive and therefore estimate the false-positive rate:

\begin{equation}
	\hat{p}_t = \frac{\left|\{s_{i,j} \in U | d_t(s_{i,j}) < \nu \}\right|}{|U|}
	\label{eq:fp}
\end{equation}

To estimate the false-negative rate, we pick one of the sample $s_{t,i}$ and calculate its average distance to the other 11 samples $s_{t,j}, j \neq i$, and count it as a false negative if the distance is at least $\nu$.  We then repeat this process for each $i=1,\ldots,12$:

\begin{equation}
	\hat{q}_t = \frac{\left|\{s_{t,i} | i \in \{1,\ldots,12 \}, \sum_{i \neq j} d(s_{t,i},s_{t,j}) / 11 \geq \nu\}\right|}{12}
	\label{eq:fn}
\end{equation}

Given a target false positive rate of $p_t^*$, we can calculate the threshold $\nu_t^*$ that would ensure $p_t < p_t^*$.  Note that
because $\hat{p_t}$  is only an estimate of $p_t$, we calculate a 95\% confidence interval for $p_t$ and chose $\nu$ such that
the upper limit of the CI is below $p_t^*$\footnote{We use a binomial proportion confidence interval here. This is slightly imprecise,
as the $12\cdot 999$ samples are not independent; we leave computation of confidence intervals that take this into account for future work.}.  Note that this threshold will be different for each site.  We can then estimate the corresponding false negative rate $\hat{q}_t^*$ that corresponds to $\nu_t^*$.

The target false positive rate will largely depend on the prior knowledge the attacker has.  Typically, we will want to aim for a small false-positive rate, since even if Bob considers it likely that Alice does in fact visit the target site $t$ \emph{at some point}, most of the web browsing in any trace will still be to other sites; thus a low false-positive rate is needed for the test to have high positive predictive value.  On the other hand, Bob can easily tolerate a moderate false-negative rate, since even if he only finds out about employees searching for other jobs 90\%, or even 50\% of the time, this information is useful nevertheless.  Likewise, perfect detection is not needed for the potential attack to have a chilling effect on Alice's behavior.

\begin{figure}
   \centering
\subfigure[VM--DSL case]{
       \label{fig:vd}
    \includegraphics[width=0.5\columnwidth]{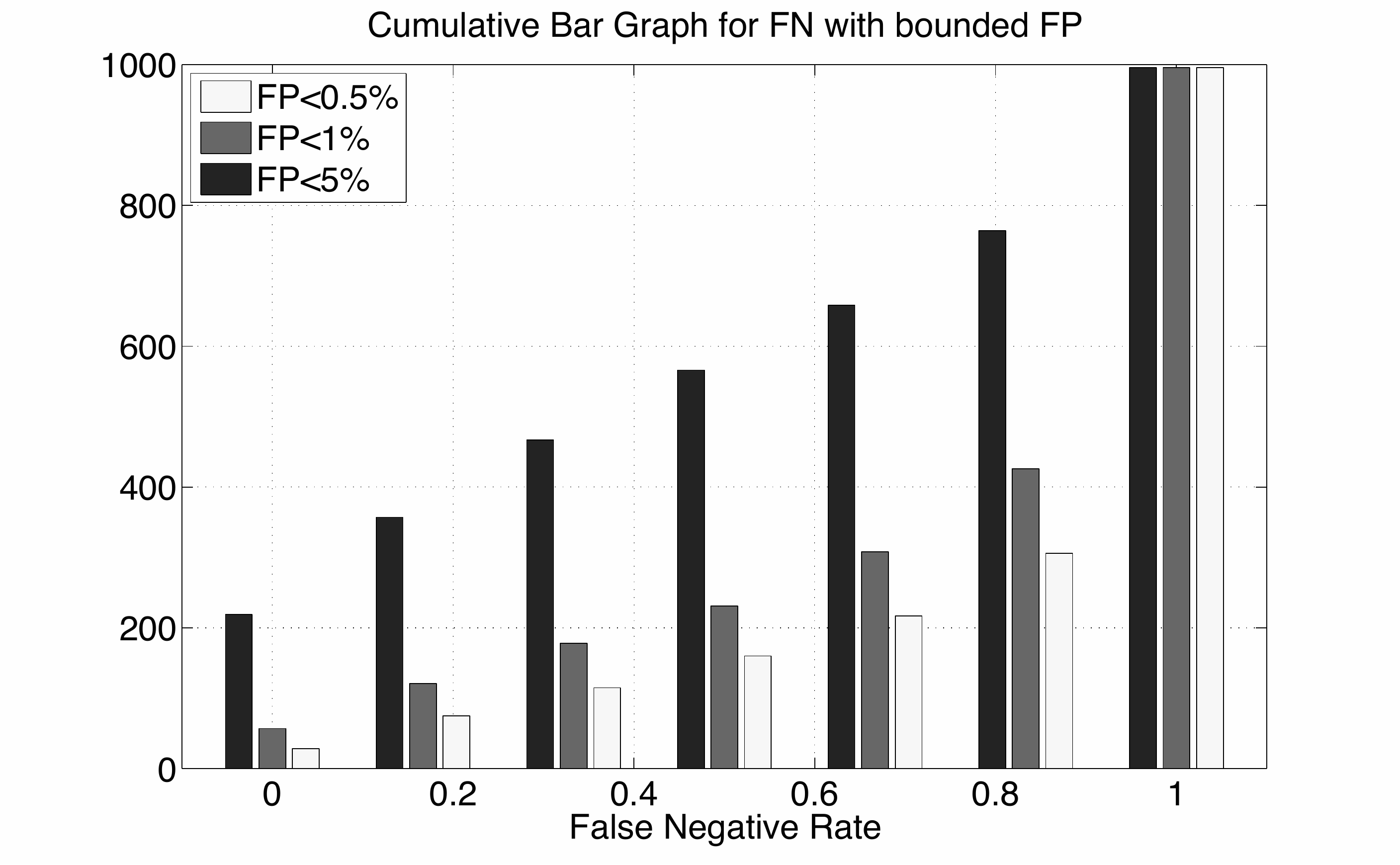}
      }\subfigure[DSL--DSL case]{
    \label{fig:dd}
    \includegraphics[width=0.5\columnwidth]{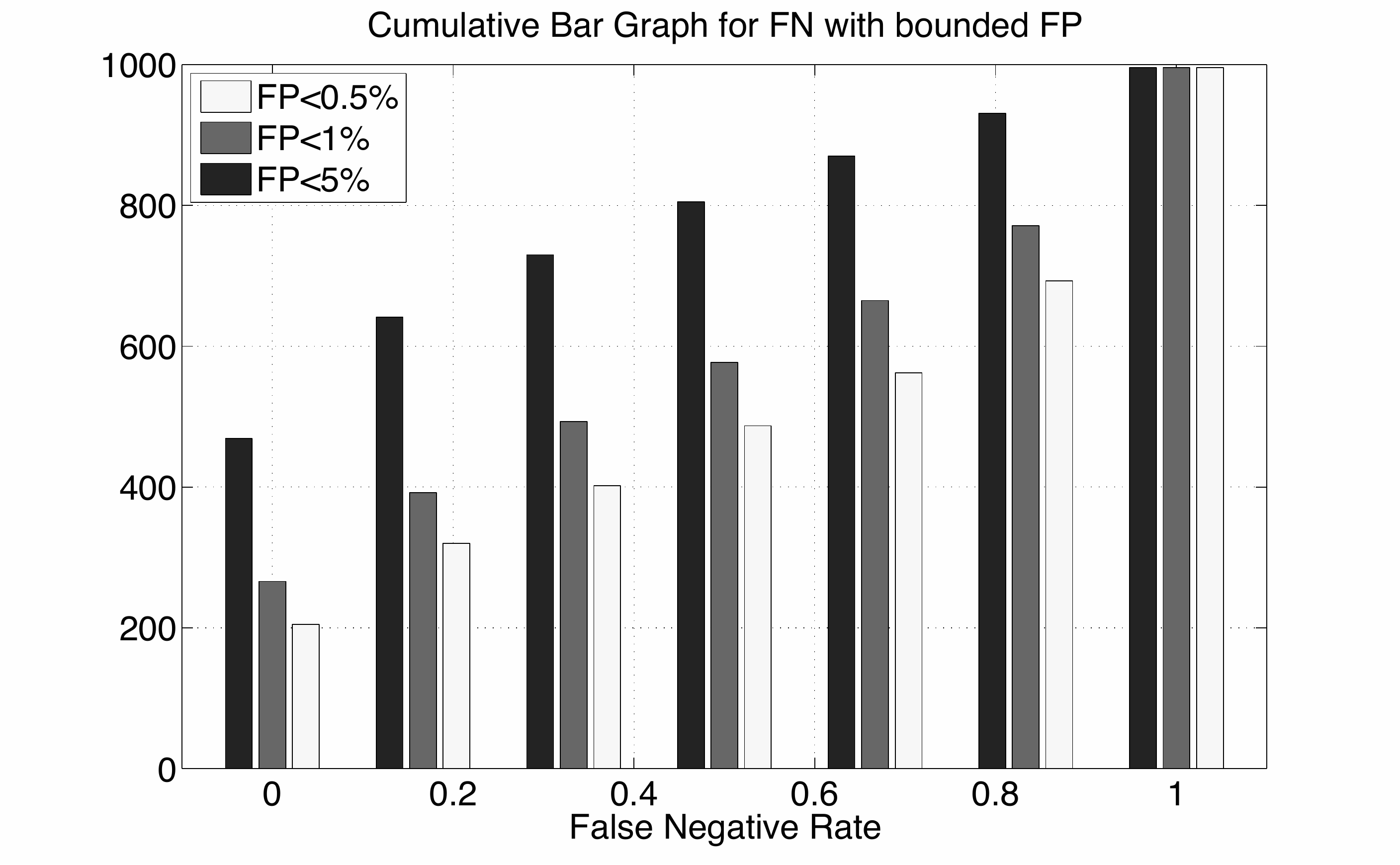}
      }  
       \caption{Number of sites with a given false negative rate or smaller.}
    \label{fig:c_fn}
\end{figure}

Figure~\ref{fig:c_fn} shows the false negative rates that can be achieved given a target false-positive rate of 0.5\%, 1\%, and 5\%.  Each
bar represents a cumulative number of websites, i.e., websites for which $\hat{q}_t^*$ is at or below the x-axis value.  We show two sets of results; one using the VM setup for training and DSL for testing (Figure~\ref{fig:vd}) and one using the DSL samples for both training and test data sets (Figure~\ref{fig:dd}).  Note that there is a significant difference between the two graphs, resulting from the discrepancies between the simulated (VM) and the test environment.  We expect that, with some work, an attacker may be able to reduce such discrepancies
by more carefully tuning the parameters of the virtual machine and the simulated link, or by using actual hardware and a real DSL line that mimics Alice's setup.  The DSL--DSL case therefore shows the limits of what can be achieved by improving the training environment.

An important observation is that, in both cases, the success of the web detection is highly dependent on the target site.  For a small number of sites---75 in the VM--DSL case and 320 in the DSL--DSL case---the web detection attack works very well: we are able to maintain a very low false-positive rate of 0.5\% while experiencing few false negatives (17\% or below).  On the other hand, some sites are virtually invulnerable to our attack: for 65 of the sites we tested, we were unable to observe \emph{any} true positives with a target false-positive rate of 5\% (i.e., $\hat{q}_t^* = 1$), even in the best-case DSL--DSL scenario.  We found these sites to have either very short traces, making it difficult to distinguish them from other such sites, or highly variable traffic patterns due, for example, to dynamic content, making it difficult to create a useful fingerprint.

\subsection{Deanonymization}

We next consider the deanonymization attack described in \S\ref{sec:attacks}.  As a case study, we considered the site \href{http://www.warriorforum.com}{www.warriorforum.com}, a popular Internet marketing forum.  It uses the vBulletin software, which was, as of August 2011, the
most popular bulletin board software\footnote{\url{http://www.big-boards.com/statistics/}}, and thus should be representative of a number of other sites.  In our attack scenario, Bob wishes to find out if Alice is using a particular pseudonym (say, ``diane123'') to post on the site.  To accomplish this,
he first collects traces from Alice's home computer for a period of time.  He then waits for posts to the forum from diane123
and performs a detection attack to see if Alice was visiting the site at the time of post.  Repeated successful matches can then be used to 
obtain increasing confidence in tying Alice to diane123.

\begin{wrapfigure}{r}{0.5\textwidth}
	\centering
	\includegraphics[width=0.5\textwidth]{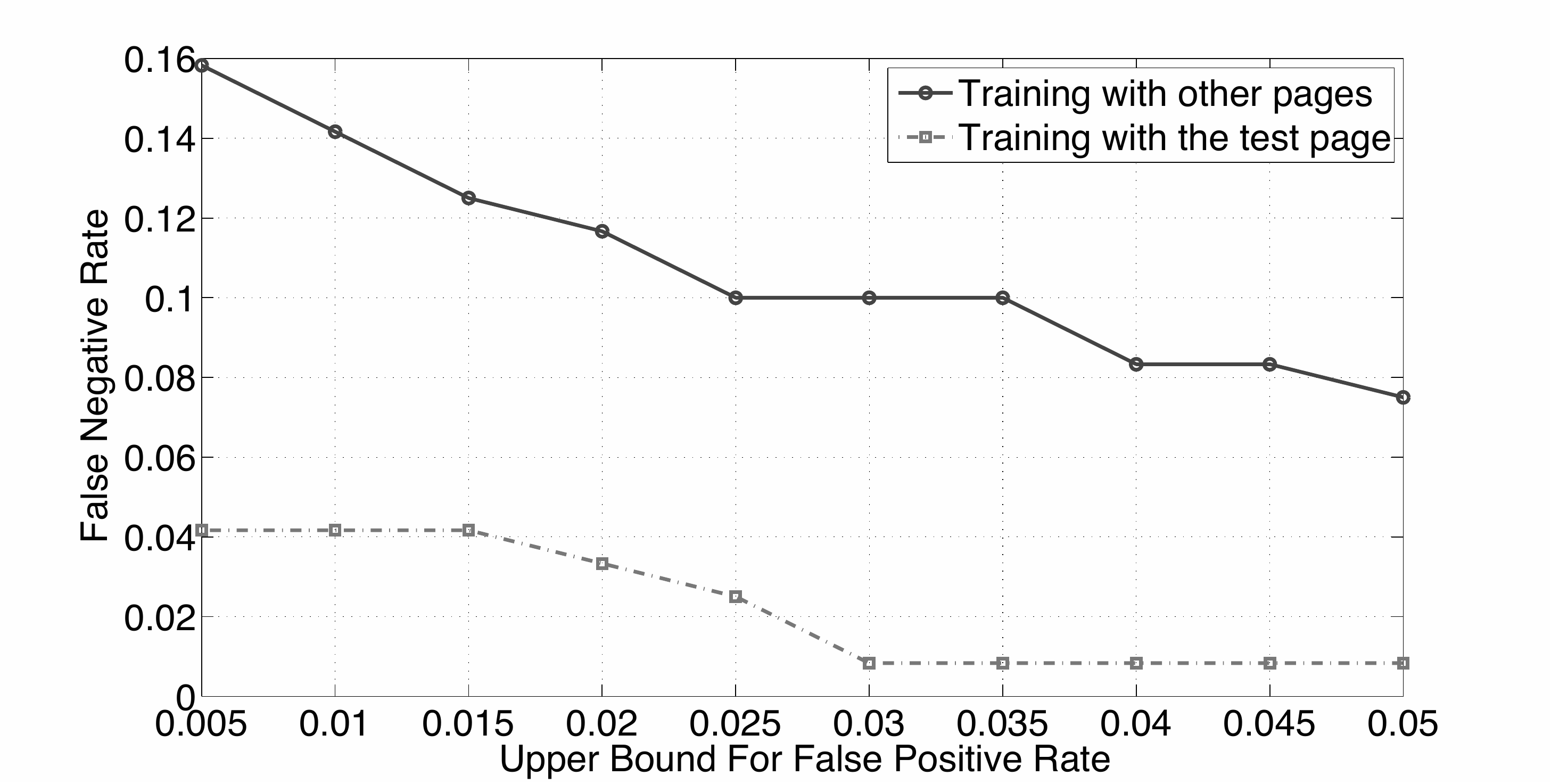}
	\caption{Detection performance for Warrior Forum when training with the correct page and when using other pages.}
	\label{fig:warrior}
\end{wrapfigure}

Note that Bob will need to build a profile that targets internal pages of \nolinkurl{www.warriorforum.com}, rather than the front page.  Alice's post requests will be too small to create an easily-observable feature; however, vBulletin displays the forum thread after a post has been made.  Therefore, Bob can collect samples visiting threads where diane123 has posted to create a fingerprint.  Note that, in this attack, fingerprint creation happens \emph{after} the trace collection.  A problem facing Bob is that different post pages on Warrior Forum will have 
similar features in their RTT profile.  A match, therefore, can show that Alice visited \emph{some} Warrior Forum page with high confidence, but it may not have been the correct thread.  Even this information, however, is likely to be enough for deanonymization.  For example, Alexa shows that fewer than 1\% of US Internet users actually visit the Warrior Forum, and those that do tend to stay on the site less than 10 minutes on average.  If we make the simplifying assumptions that these visits are distributed randomly across a three-hour evening period, the additional false-positive rate due to random visits to \emph{other} Warrior Forum pages is no more than 6\%, even if Alice is \emph{known} to be a Warrior Forum user.  Combining observations across several posts allows Bob to improve his confidence.

Over a long term, even simpler attacks may suffice: since most people's Internet usage is bursty, simply observing that Alice always actively used the Internet in some way whenever diane123 made posts can be used for deanonymization~\cite{danezis2005statistical}.  Likewise,
Bob may be able to rule out Alice as a suspect if she was known to be at home (due to recent DSL activity) but her connection was idle at the times of the target posts.

Finally, Bob may be able to use the similarity between internal forum pages to his advantage.  In particular, suppose that Alice publicly participates in the forum under her real identity, in addition to potentially posting under a pseudonym.  Bob can use the times of Alice's posts under her real name to label traces collected from Alice's computer and create a training set.  In this case, Bob does not need to simulate Alice's computing environment as the training and test environments are exactly the same---the ideal conditions we used 
in the DSL--DSL case.  To study this attack, we collected samples from 100 different posts on the Warrior Forum site.  For each sample, we attempt to match it to a fingerprint created from the other 99 posts; our process is similar to \eqref{eq:fn}, except using a different for each sample.  From this, we estimate the false negative rate for a given target false-positive rate, calculated using  \eqref{eq:fp}, using the traces from 999 other websites.  Figure~\ref{fig:warrior} shows the results.  The use of different pages to test degrades the matching performance, but it still provides sufficient detection power for deanonymization after a few posts.  


\section{Discussion}
\label{sec:discussion}

In this section, we discuss about  some limitations of our work. 

\begin{enumerate*}
\item \emph{Multiple users.} 
 In  the scenario of our remote traffic analysis,  the attacker's probes cannot distinguish between the traffic of multiple users on the same link, so shared broadband connections present an obstacle to our attack. However, even in multi-user installations, it is still common for only one user to be using the Internet at any given point during the day.  Some previous work on  traffic analysis has used blind source separation to separate traffic from multiple users~\cite{zhu:pet05}; similar techniques may be applicable here.  For example, in Figure~\ref{fig:trafficvsrtt}, traffic follows a periodic pattern based on the RTT between Alice and the website; such periodicity might help separate the sources.

\item \emph{Dynamic nature of websites.} Our attack relies on web sites having relatively stable fingerprints.  
Although the overall pattern captured by our RTT probes remains static enough within days, the website content may incur significant changes (e.g., site redesigns) over time; which in turn will result in a change of its fingerprint. Thus, for best results, the training set should be updated continuously.  This limitation applies to any website fingerprinting approach even local website fingerprinting techniques which benefit from better vantage points~\cite{liberatore-levine:ccs06,Herrmann2009}. 

\item \emph{Content distribution networks.}
Websites that use content distribution networks (CDNs) will use different servers to deliver content based on the
user's location.  They may present localized versions of the site to users in different countries or regions.
As shown in our experimental results, this can cause fingerprints to differ significantly.  If identifying these sites
is a high priority for the attacker, additional work would be needed to obtain fingerprints of the right version by, for example,
using proxies and other techniques to fool IP-based localization.

\item \emph{Cache issues}  In our tests, we followed the assumption in previous work~\cite{liberatore-levine:ccs06,Herrmann2009} and disabled the cache in the browser. This implies that our results demonstrates the attacker's ability to verify that a user visits a web page for the first time. 
To investigate cases with cache enabled, one possible solution would be build separate fingerprints based on the time since the site was first downloaded, e.g., after 1 hour, 6 hours, 1 day, 1 week, to minimize the effect that caching would have on the attack.  Note that with
continuous observation of a computer, the attacker may be able to guess how long ago the last visit was.

\end{enumerate*}

\section{Related Work}
\label{sec:related}

The use of network probes to infer information about traffic at a remote location has been explored in previous work in the context of anonymous communication networks.  Murdoch and Danezis used a remote traffic analysis approach to expose the identity of relays participating in a circuit in the Tor~\cite{Dingledine2004} and MorphMix~\cite{morphmix} anonymous communication networks~\cite{Murdoch2005}.
Their approach was to send an on--off pattern of high-volume traffic through the anonymous tunnel and a low-volume probe to a router under test.  If the waiting times of the probe showed a corresponding increase during the ``on'' periods, the router was assumed to be routing the flow. However, when Murdoch and Danezis evaluated their attack, the Tor network was lightly loaded and consisted only of 13 relays; to repeat their attack on today's network, with around 2\,000 relays and high traffic load\footnote{See \url{http://torstatus.blutmagie.de} (retrieved November 2010)},
an attacker would needs extremely large amounts of bandwidth to measure enough relays during the attack window. Evans et al.~\cite{evans+:sec09} strengthened Murdoch and Danezis's attack of  by a bandwidth amplification attack which make their attack feasible in modern-day deployment of Tor.   Hopper et al.~\cite{hopper+:ccs07,hopper+:tissec10} use a combination of Murdoch and Danezis's approach and pairwise round trip times (RTTs) between Internet nodes to correlate Tor nodes to likely clients. Chakravarty et al.~\cite{chakra} propose an attack for exposing Tor relays participating in a circuit of interest by modulating the bandwidth of an anonymous connection and then using available bandwidth estimation to observe this pattern as it propagates through the Tor network.  Note that these techniques relied on detecting a specially-crafted coarse-grained communication pattern, whereas our attacks make use of fine-grained information obtained through remote traffic analysis.

\label{sec:survey}  We also survey previous work on recovering information about encrypted HTTP traffic.  The fact that object sizes could be used to 
infer sensitive information, even after encryption, was first mentioned by Yee (as related by Wagner and Schneier~\cite{wagner-schneier}).  A specific concern listed by Yee is that the particular page within a site accessed by the user could be revealed by considering URL and object lengths.  Chen et al.~\cite{chen+:oakland10} applied this observation to AJAX applications to recover detailed information about the internal state of the application and users' data.  

Cheng et al.\ \cite{Chen98trafficanalysis} present the earliest implementation of website fingerprinting. The classification features used in their scheme are the object sizes and the HTML file sizes.  
Hintz \cite{Hintz02} and Sun et al.\ \cite{Sun02} both consider website fingerprinting attacks in SSL-encrypted HTTP connections.
Their classification features are object sizes and counts. While Hintz did not present implementation details and experimental results,
Sun et al.\ use a Jaccard's coefficient based classifier and show that their attack can achieve a correct identification rate of 75\%.

Instead of looking at web objects,  Bissias et al.\ \cite{bissias+:pet05}, Liberatore et al.\ \cite{liberatore-levine:ccs06},  and Herrmann et al.\ \cite{Herrmann2009} study the statistical characteristics of individual packets in the traffic flows. Bissias et al.\  use packet sizes and inter-arrival timings as classification features. Their method is fragile to the changes in the network environment, as the inter-arrival timing is highly dependent on the specific routing path and varies
from time to time. To address this problem, Liberatore et al.\  only use packet sizes and counts in classification. They implement both Jaccard coefficient and Na{\"i}ve Bayes classifier, and show the efficacy of the attack in practice. Using similar scheme,  Herrmann et al.\ further improve the classification accuracy using Multinomial Na{\"i}ve Bayes classifier.

\section{Conclusion}
\label{sec:con}
We show that traffic analysis attacks can be carried out remotely, without access to the analyzed
traffic, thus greatly increasing the attack surface and lowering the barrier to entry for conducting the attack.
We identify a queuing side channel that can be used to infer
the queue size of a given link with good accuracy and thus monitor traffic patterns.  We show how this channel can
be used to carry out a remote attack to detect a remote user's browsing patterns.  This
highlights the importance of traffic analysis attacks in today's connected Internet.

\bibliographystyle{nikita-tr}
\bibliography{nikita}

\end{document}